# Time-domain terahertz compressive imaging

L. Zanotto[1], R. Piccoli[1,3], J. Dong[1], D. Caraffini[1], R. Morandotti[1,2] and L. Razzari[1,4]

[1]*Institut National de la Recherche Scientifique (INRS), Centre Énergie, Matériaux et Télécommunications (EMT), Varennes, QC J3X 1S2, Canada.*
[2]*Institute of Fundamental and Frontier Sciences, University of Electronic Science and Technology of China, Chengdu 610054, China.*
[3]*riccardo.piccoli@emt.inrs.ca,*
[4]*razzari@emt.inrs.ca*

**Abstract:** We present an implementation of the single-pixel imaging approach into a terahertz (THz) time-domain spectroscopy (TDS) system. We demonstrate the indirect coherent reconstruction of THz temporal waveforms at each spatial position of an object, without the need of mechanical raster-scanning. First, we exploit such temporal information to realize (far-field) time-of-flight images. In addition, as a proof of concept, we apply a typical compressive sensing algorithm to demonstrate image reconstruction with less than 50% of the total required measurements. Finally, the access to frequency domain is also demonstrated by reconstructing spectral images of an object featuring an absorption line in the THz range. The combination of single-pixel imaging with compressive sensing algorithms allows to reduce both complexity and acquisition time of current THz-TDS imaging systems.

## 1. Introduction

The possibility of performing imaging at terahertz (THz) frequencies (0.1-10 THz) has received a continuously growing attention in the last decades [1]. The primary reasons are the unique ability of THz radiation of "seeing through" a multitude of optically-opaque dielectric materials such as plastics, semiconductors, papers [2], as well as the low THz photon energy (meV instead of keV associated with X-rays) that makes it intrinsically safe to use with sensitive and biological materials. Moreover, the capability of performing time-domain coherent detection (amplitude and phase information) of the THz electric-field waveform enables full material characterization, including the retrieval of the complex refractive index [3] and the chemical recognition of substances (e.g., gases [4], organic materials [5], bio-molecules [6]) that feature rotational/vibrational transitions, or collective excitations (e.g., phonons and magnons) in the THz range. This make THz imaging a promising technology for a multitude of applications including security [7], quality and safety control in industry [8,9], biomedical applications [10,11] and also art conservation [12,13]. The main hurdle that still hampers wide spread application of THz imaging technology is the long acquisition time [14]. Indeed, the lack of affordable THz cameras leads most common systems to rely on raster-scanning to retrieve a THz image via single-pixel detectors [15]. Moreover, one of the most powerful techniques employed in THz science, THz time-domain spectroscopy (TDS), typically makes use of detectors with no spatial resolution, such as photo-conductive antennas and electro-optic crystals. In this popular case, an object has to be raster-scanned pixel-by-pixel and, for each one of them, an entire temporal scan is required to reconstruct the THz temporal waveform, thus dramatically increasing the acquisition time. Therefore, an imaging scheme able to address the long acquisition drawback, whilst maintaining the advantages of TDS detection, would be of great interest to practically exploits the capabilities of this technology in real-world environments. Numerous solutions have been proposed to address the long acquisition time issue, ranging from THz cameras based on micro-bolometers [16] (with no phase/spectral resolution) and fast mechanical beam steering [17] to various computational imaging implementations [18–20]. A promising paradigm is the so-called single-pixel imaging scheme.

It allows the indirect reconstruction of the object by illuminating it with a series of spatial patterns and, for each of them, recording the total transmission/reflection using a detector with no spatial resolution (i.e., "single-pixel") [21]. In this way, raster-scanning of the object is avoided and, in addition, compressive sensing (CS) algorithms can be implemented, which allow to recover a good approximation of the image with a number of measurements significantly smaller than the number of pixels [22]. CS exploits the fact that the images of most objects are sparse under certain representations (i.e., most of the elements of the image are zero). Thus, with a proper choice of the illuminating patterns basis, one can reconstruct the image with an "incomplete" set of measurements, resulting in a significant reduction of the acquisition time. The first attempt to employ a single-pixel configuration for THz imaging, applied a series of metallic masks in order to encode the patterns onto the THz beam [18,23]. Following this path, in the last few years, research efforts were directed towards finding more effective ways to spatially modulate the THz beam, for example via optical photo-excitation of semiconductors [24,25], or with metamaterial-based spatial light modulators [26]. Lately, exploiting the flexibility of the optical modulation technique, several studies focused on demonstrating the use of single-pixel techniques for *near-field* imaging [27,28], including a very recent proposition also featuring the time-domain reconstruction at different spatial positions [29]. In this work, we present a *far-field* imaging implementation based on the use of single-pixel imaging with a THz-TDS system. By indirectly reconstructing, for each individual pixel, the THz waveforms in the time-domain, we are able to retrieve images in multiple dimensions (space/time/frequency). First, we exploit the temporal information available in each pixel to realize (for the first time in THz single-pixel imaging, to the best of our knowledge) time-of-flight images of a selectively-carved High-Density Polyethylene (HDPE) sample, made via 3D printing. Moreover, as a proof-of-concept, we apply a typical compressive sensing algorithm to demonstrate image reconstruction with less than 50% of the total number of measurements (i.e., THz spatial patterns). Finally, the access to the frequency domain via Fourier transformation is also reported, by recording spectral images of an object composed of sections made of Teflon and of a Teflon/lactose mixture (lactose featuring a clear absorption line at 1.3 THz [30]). We believe that our approach paves the way towards fast multidimensional imaging at THz frequencies, which can be directly implemented in traditional TDS setups.

## 2. Experimental implementation

### 2.1 Single-pixel imaging

The idea behind single-pixel imaging is to interrogate the object with a series of spatial patterns, encoded onto the beam used for imaging. Typically, the total intensity of the radiation from each of these patterns is recorded by a single-pixel detector, after the interaction with the object. The correlation between the spatial shape of each pattern and the measurement from the detector enables the reconstruction of the image. Considering just transmission for simplicity, the $n \times n$ sample transmission function $T(x,y)$ can be seen as a $n^2$-element vector $S$, by placing each row of the matrix one next to the other. The collection of measurements from the detector corresponds to another $n^2$-element vector, $W$. Finally, the series of $n^2$ patterns $\{H_i(x,y)\}_{i=0...n^2}$ can be reshaped into $n^2$ basis-vectors with $n^2$ elements, by concatenating the rows of each pattern and grouping them into a $n^2 \times n^2$ matrix $H$. In this fashion, the measurement process can be simply seen as a matrix-vector multiplication [28]. Thus, the knowledge of the matrix $H$ and the vector $W$ allows the retrieval of the object shape $S$, by solving a linear system of equations:

$$W = HS \qquad (1)$$

The application of a single-pixel imaging scheme into a THz-TDS system requires the use of an electric-field-sensitive detector. This means that the detector, in this case, measures an electric field as a function of time ($t$), with an amplitude that can be positive or negative at each

*t*. We can thus rewrite Eq.1, substituting vectors *W* and *S* with their time-dependent versions *W(t)* and *S(t)*, representing the electric field output of the detector and the electric field waveform retrieved at each pixel position, respectively:

$$W(t) = \mathbf{H}S(t) \qquad (2)$$

Solving the linear system of Eq.2, we can reconstruct the THz electric field as a function of time at each spatial position of the scene, therefore adding a third dimension (time) to the two spatial dimensions.

## 2.2 Compressive sensing implementation

As mentioned above, one of the advantages of single-pixel imaging is the possibility to apply CS algorithms. This allows for a much faster recovery of the object under investigation by using an incomplete set of measurements. In principle, in order to perform the matrix inversion of Eq.1, one needs the dimension of the vector *W* to be equal to that of the vector *S*, which means performing $n^2$ measurements. Nonetheless, the CS theory shows that, under particular representations, it is possible to recover the shape of the object *S* with good approximation by performing a number *m* of measurements smaller than the number of pixels ($m < n^2$), and applying specific algorithms to best exploit the object sparsity [22]. There is a variety of possible choices of pattern bases to implement the single-pixel imaging scheme. One of the most commonly used is the one derived from the Hadamard matrix, which has been proved to be good in signal-to-noise ratio (SNR) maximization [27]. In this work, we made use of an optimized ordering of the Hadamard set, as proposed by M.J. Sun et al. [31], in which the spatial frequencies content of each pattern increases progressively. By using this scheme, most of the information related to the image will be contained in the first patterns. This allows to decrease the acquisition time (by reducing the number of measurements required to reconstruct the image), while avoiding the calculation burden usually necessary for compressive sensing schemes based on random pattern sets [31]. For the image reconstruction, we made use of a non-iterative single-pixel algorithm, as reported by L. Bian et al. [32].

## 2.3 Experimental setup

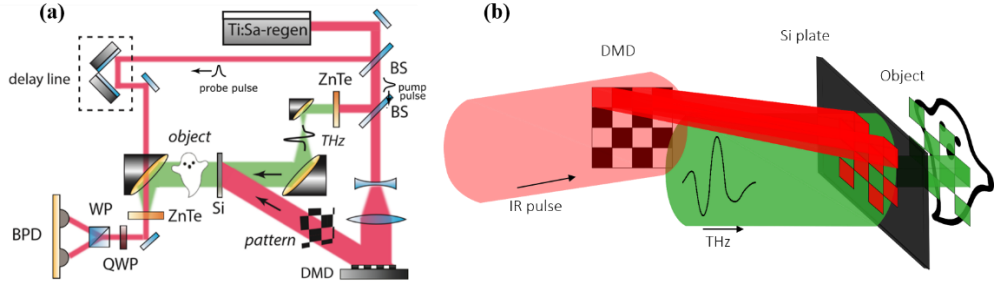

Figure 1: (a) Experimental setup: modified version of a THz-TDS system. The beam from the fs-laser is split into three lines, one generates the THz pulse, one is used as probe in the electro-optic sampling crystal and the last one is employed for THz modulation. (b) THz modulation technique: spatial patterns are encoded onto the infrared laser beam by means of a DMD; the patterned beam illuminates a Si-plate where it generates photo-carriers. The THz beam passing through the Si, right after photo-excitation, is locally reflected and thus acquires the same spatial pattern of the infrared laser beam.

The experimental setup is a modified version of a traditional THz-TDS system (Fig.1(a)). An amplified Ti:Sapphire laser (800 nm, 150 fs pulse duration, 2 mJ at 1 kHz repetition rate) is split into three beams: pump (48% of total laser power), probe (4%) and pattern generator (48%). The pump beam is used to produce THz pulses via optical rectification in a 1-mm-thick zinc telluride (ZnTe) crystal. The THz beam is first enlarged by means of a series of gold off-axis parabolic mirrors, propagates through a 500-μm-thick silicon plate (undoped, 5000 Ω·cm) used for spatial modulation, and then passes through the sample. Finally, the THz waveform is recorded via electro-optic sampling in another 1-mm-thick ZnTe crystal, by means of the probe beam. To implement a single-pixel imaging technique at THz frequencies, THz spatial patterns are required for interrogating the sample. Commonly used spatial light modulators are not suitable for operation in the THz range. We spatially modulate the infrared pattern generation beam by means of a digital micro-mirror device (DMD) (*LightCrafter4500*, Texas Instruments) and illuminate the surface of the silicon plate (Fig.1(b)) with the modulated beam. Silicon is normally transparent in the THz range, but under laser illumination (with photon energy above its bandgap) free carriers are locally generated, thus preventing the transmission of THz light and transferring the pattern spatial distribution from the infrared laser to the THz beam. By applying this technique, we achieve a modulation depth of about 95% (energy of the THz beam) with an infrared pulse energy of ~140 μJ incident on the silicon plate over an area of $1.6 \times 1$ cm$^2$.

## 3. Results and discussion

The reconstructed THz waveforms obtained via the implemented single-pixel imaging scheme are presented in Fig.2. On the left, we show the measured THz waveforms acquired while illuminating with some selected patterns a sample made of HDPE (rectangular shape 1.6 cm × 1 cm and 2 mm thick), with "TZ" letters carved (where "T" is 2 mm deep and "Z" is 1 mm deep, as measured from the sample surface). We can see a series of THz pulses, due to the contributions from THz waves that passed through different positions on the sample, therefore arriving at different times. On the right side of Fig.2 we have the THz traces reconstructed using Eq.2 for three representative pixels (for "T", "Z" and "background" positions). The ability to retrieve the temporal waveform at each spatial position gives access to the full amount of information one would obtain by raster-scanning the object, allowing to perform imaging of objects with different contrast mechanisms. This evidence demonstrates the capability of this imaging configuration to yield multidimensional far-field images of the object.

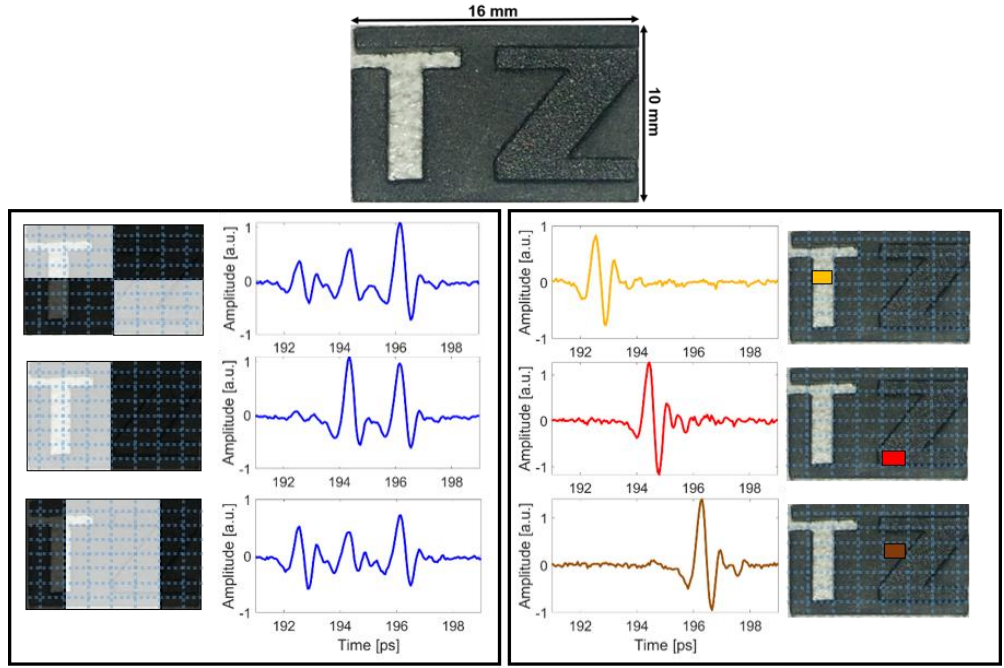

Figure 2: (top) HDPE sample with letters "T" 2 mm deep and "Z" 1 mm deep. (left) detector output for three of the spatial patterns used. The THz waveforms are multi-pulse, due to THz waves arriving at different times after passing through the sample at different positions. (right) THz electric field reconstructed in time at three selected pixels.

### 3.1 Time-of-flight images

One implementation that exploits the full potential of THz-TDS systems relies on the retrieval of the phase-delays at various spatial positions to reconstruct the thickness (and/or the density/refractive index) of a sample. This technique is extremely powerful when dealing with materials transparent to THz radiation, for which the contrast based on direct amplitude measurements may be low. In Fig.3(a) we can see the time-of-flight reconstruction of a 256-pixel image of the "TZ" sample made of HDPE (refractive index n = 1.58 at 1 THz). Reconstruction is carried out considering the relative delay of THz pulses at each pixel position. From these delays, it is possible to extract the actual thickness of the sample, knowing the refractive index of the material, therefore achieving a real three-dimensional imaging. In this regard, we can also envision the possibility of applying deconvolution algorithms [33], after the temporal traces are retrieved, to resolve subwavelength multilayered structures. We subsequently used the same sample to test the effectiveness of compressive sensing reconstruction in the context of our direct temporal THz single-pixel imaging configuration. The results are shown in Fig.3(b), where we can see how reducing the number of measurements results in losing some details of the image, although the shape of the letters is still recognizable even with only 40% of the total number of patterns, and it is still possible to extract the thickness at each spatial position.

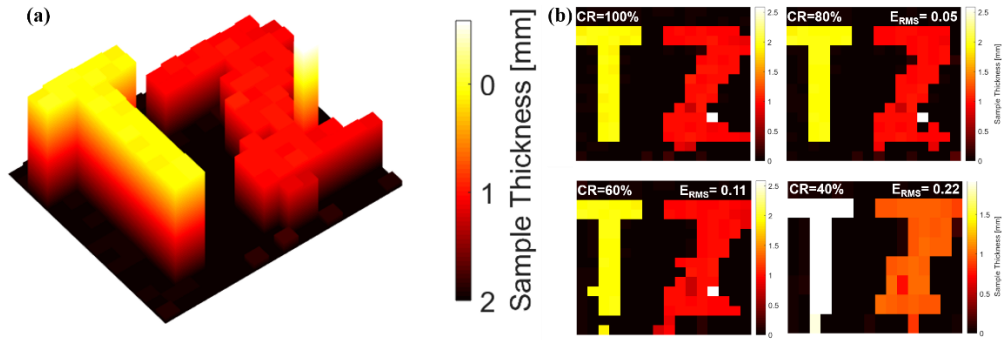

Figure 3: Time-of-flight image: (a) 256-pixel image reconstructed using the local thickness of the HDPE sample described in the text. The thickness is retrieved with the relative time delay of the THz pulses at each spatial position. (b) Image with variable compressive ratio (CR, number of measurements/number of pixels): from top-left CR=100%, CR=80%, CR=60% and CR=40%. The Root Mean Squared Error ($E_{RMS}$) is calculated averaging the error at the THz wave peak over all the pixels in the image.

### 3.2 Hyper-spectral images

Another key application of such imaging scheme deals with the spectroscopic recognition of the chemicals present in a sample. Indeed, the THz transient retrieved in each pixel carries full information about the sample, including absorption lines at specific frequencies characteristic of certain substances. To prove the possibility of spectroscopic recognition via imaging using our technique, we realized the sample shown in Fig.4(a), made of Teflon (PTFE, refractive index ~1.45 at 1 THz and no absorption) and of Teflon/Lactose mixture sections of equal thicknesses (0.7 mm). We chose Lactose due to its absorption line at ~1.3 THz [30], which can be seen in Fig. 4(d) in the spectra (retrieved via direct Fourier transformation of the reconstructed THz waveforms) of some selected pixels. The reconstruction of 8×8-pixel spectral images is made by post-selecting the transmission in correspondence of the Lactose absorption line at ~1.3 THz (Fig.4(c)) or at 1 THz (Fig.4(b)). We can see that the L-shaped part made of the Teflon/Lactose mixture results darker overall in Fig.4(c), while the rectangle made of PTFE shows not significant difference from the rest of the imaging window. The differences in absorption (color bar) among different pixels within the L-shaped part containing Lactose are probably due to its inhomogeneous spatial distribution within the mixture. No features are instead visible for the spectral image at 1 THz (Fig.4(b)).

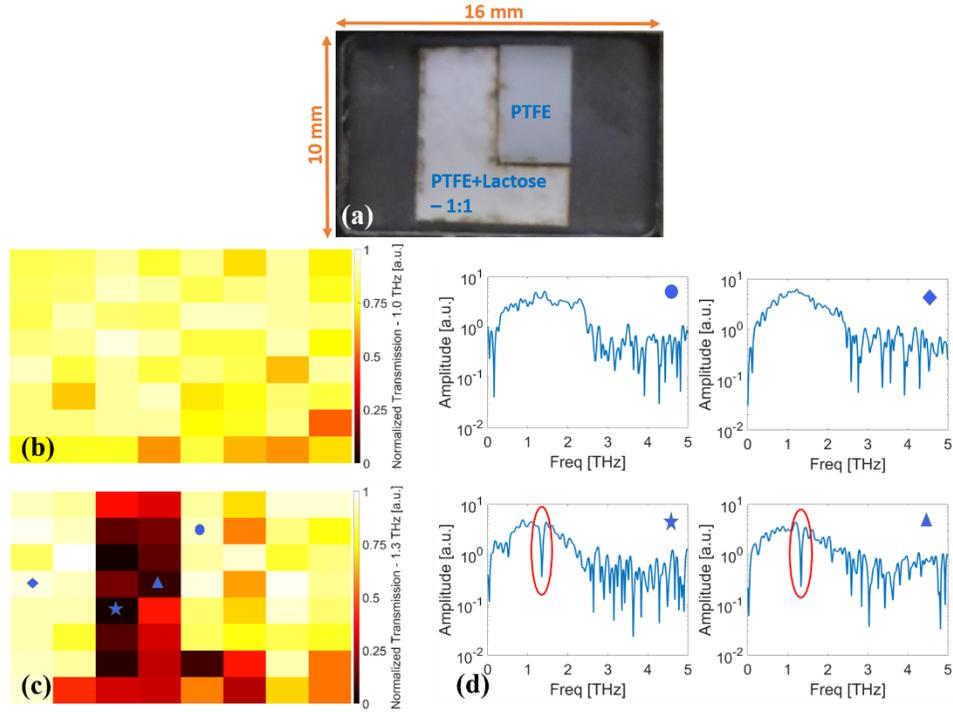

Figure 3: Hyper-spectral images: (a) sample: L-shaped pellet made of a 1:1 mixture of Lactose and PTFE (index of refraction 1.45 at 1 THz) and a rectangular pellet made of PTFE. (b) 64-pixel spectral image at 1.0 THz. (c) 64-pixel spectral image at 1.3 THz (d) example of THz spectra of four selected pixels, obtained by direct Fourier-transformation of the time-domain THz waveforms reconstructed at each pixel position: for the pixels with Lactose the absorption line at 1.3 THz is clearly visible.

## 4. Conclusions

Summarizing, we have applied a single-pixel imaging configuration to a standard THz-TDS system. We have demonstrated the reconstruction of the THz electric field waveform in the time-domain at each spatial position of the object, without the need of mechanical raster-scanning. We have exploited this capability to retrieve time-of-flight images of a sample made of HDPE with variable thickness, and hyper-spectral images of a sample containing Lactose (featuring an absorption line at 1.3 THz) and Teflon. This demonstrates the flexibility of the technique to retrieve far-field images in multiple dimensions. Moreover, we have shown that CS algorithms can be easily implemented, thus leading to a significant reduction of the overall acquisition time. We believe that this work, combining the powerful capabilities of THz-TDS and CS, is of relevance for the next generation of THz imaging technologies featuring faster acquisition times.

## 5. Funding

Natural Sciences and Engineering Research Council of Canada (NSERC) (Strategic Grant: *Terahertz hyper-imaging systems for advanced manufacturing*); Prima Quebec; Fonds de Recherche du Québec - Nature et Technologies (FRQNT); MITACS.

## Acknowledgments

We are indebted to our industrial partners TeTechS, Lumenera, and Integrity Testing Laboratory (ITL) for the help offered at various stages of this project. Note that R.M. is affiliated to (2) as adjoint faculty.